\documentclass[openacc]{rstransa}%%%%where rstrans is the template name

\def\nustar{{\em NuSTAR}}
\def\chandra{{\em Chandra}}

\def\cs{$C_{\rm S}$}
\def\cb{$C_{\rm B}$}
\def\snmin{SN$_{\rm min}$}
\def\Dt{$\Delta T$}
\def\farcs{\hbox{$.\!\!^{\prime\prime}$}}
\newcommand{\angstrom}{\mbox{\normalfont\AA}}
\def\fdeg{\hbox{$.\!\!^{\circ}$}}

\newcommand{\revised}[1]{{#1}}
\titlehead{Opinion}

\begin{document}

%%%% Article title to be placed here
\title{X-ray astronomy from the lunar surface}

\author{%%%% Author details
P. Gandhi$^{1}$}

%%%%%%%%% Insert author address here
\address{$^{1}$School of Physics \& Astronomy, University of Southampton, SO17\,1BJ, UK}

%%%% Subject entries to be placed here %%%%
\subject{X-ray astronomy, Moon}

%%%% Keyword entries to be placed here %%%%

%%%% Insert corresponding author and its email address}
\corres{P. Gandhi\\
\email{poshak.gandhi@soton.ac.uk}}

%%%% Abstract text to be placed here %%%%%%%%%%%%
\begin{abstract}
Motivated by efforts to return humanity to the Moon, three cases are reviewed for X-ray astronomy from the lunar surface: (1) Facilitation of ambitious engineering designs including high throughput telescopes, long focal length optics and X-ray interferometery; (2) Occultation studies and the gain they enable in astrometric precision; (3) Multimessenger time-domain coordinated observations. The potential benefits of, and challenges presented by, operating from the Moon are discussed. Some of these cases have relatively low mass budgets and could be conducted as early pathfinders, while others are more ambitious and will likely need to await improvements in technology or well-developed lunar bases. 
\end{abstract}
%%%%%%%%%%%%%%%%%%%%%%%%%%%

%%%%%%%%%% Insert the texts which can accomdate on firstpage in the tag "fmtext" %%%%%

\begin{fmtext}
\section{Introduction}
%%%% Insert A head here

X-ray astronomy is just over six decades old, and has been fundamental to unlocking the hot and energetic universe on all scales \cite{giacconi62}. Detection of high energy cosmic photons can only be done from beyond the Earth's atmosphere. Focusing of X-rays also requires specialised designs different from the normal incidence telescopes in wide use at lower energies \revised{(e.g. in the optical or infrared)}. This has posed important constraints on the development of the field so far. 

For example, grazing photon incidence is the primary mode employed in X-ray imaging optics. At the shallow angles necessary for bringing high energy X-rays \revised{reflecting off a mirror} to a focus, the requisite focal lengths start to become impracticably long for single satellites in orbit. Furthermore, deployment and control of large area telescopes in orbit is not currently possible due to payload launch constraints. As a result, X-ray focusing telescopes are restricted to effective collecting areas of order a few hundred sq.\,cm, far smaller than telescope mirrors commonly available at lower energies. 

Unlike the Earth, lunar X-ray astronomy could be conducted from the surface. Telescopes and their supporting infrastructure would have to contend with

\end{fmtext}

%%%%%%%%%%%%%%% End of first page %%%%%%%%%%%%%%%%%%%%%

\maketitle

\noindent
surface gravity, but one that is low relative to Earth. This could substantially ease, and even circumvent, the above engineering challenges, and also open up other novel directions of scientific developments, though not without introducing new challenges.

Humanity's return to the Moon with the Artemis missions and proposed lunar settlement plans currently being formulated (cf. Carpenter, this volume) provide the opportunity to explore new science cases that may benefit from operation on the lunar surface. Here, we outline three such cases for X-ray astronomy. We discuss how operating from the lunar surface could facilitate the implementation of complex engineering design concepts, including large area and long X-ray focal length telescopes, as well as multi-telescope interferometry. Following this, two science cases are developed, centred on occultation studies of bright cosmic targets in order to enable X-ray astrometric studies, and a case for coherent multiwavelength approaches to coordinating simultaneous observations. \revised{We end with a discussion of several key technical challenges that will need to be overcome when operating X-ray facilities on the Moon, and relevant considerations for choosing appropriate observatory sites. 

There have been previous proposals to study the Solar wind and Earth's magnetosheath from the vantage point of X-ray observatories on the lunar surface \cite{lxo}, but here we go beyond to develop wider cases for general X-ray astronomical studies.} Similar ambitions existed as early as 1990 \cite{gorenstein90}, though the progress of the field has proven to be more challenging than anticipated at that time. This article is not meant as an exhaustive review. Instead, it should be read as a pilot proposal on the need to consider and develop novel scientific exploitation opportunities offered by our nearest cosmic neighbour, even for the kinds of science that we have been carrying out from beyond the Earth for decades. 

\section{Facilitating Ambitious Engineering Designs}

\begin{figure}
    \centering
    \includegraphics[angle=0,width=\textwidth]{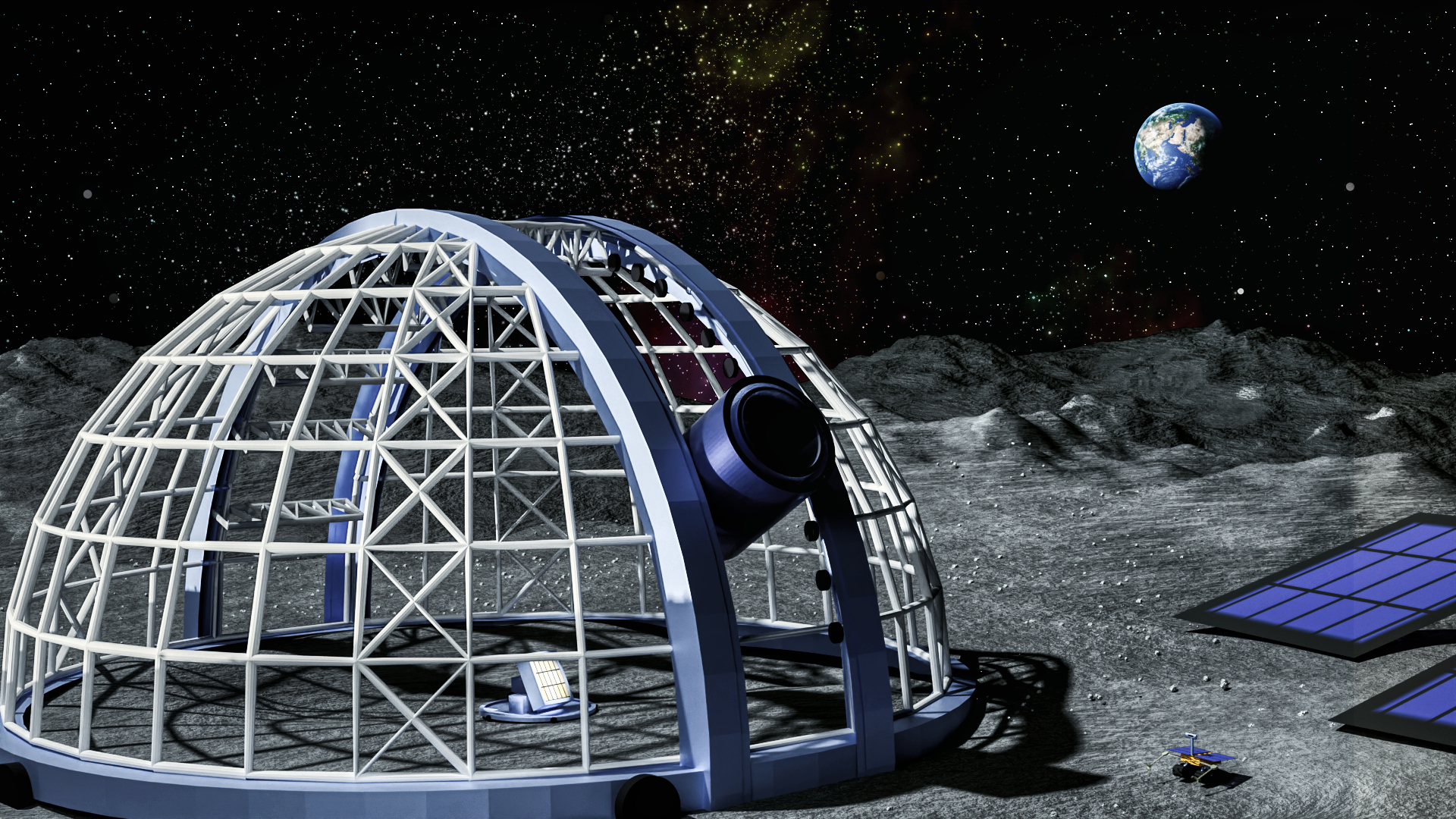}
    \caption{Artist's impression of a focusing X-ray telescope in a dome on the lunar surface. The dome is a mesh structure supporting and directing the mirrors. X-ray detectors are situated at the centre of the dome on the ground, separate from the mirror structure but tracking-locked. Additional collimators and detector shielding material will be required, but has been stripped away for clarity in this impression. Lunar topography and power units are visible in the background, with much of the control electronics expected to be buried below the surface regolith.  An imagined autonomous servicing rover is depicted in the foreground. Credit: S. Mandhai inc.  @TheAstroPhoenix.}
    \label{fig:xrayobs}
\end{figure}

\subsection{Larger Area, Longer Focal Length Optics}

Focusing X-ray photons requires grazing incidence optics, with the current dominant design being the Wolter type \cite{wolter52}. The relevant critical  angles ($\theta$) are shallow, and scale inversely with photon energy $E$: 

\begin{eqnarray}
    \theta &=& 5.6\, {\lambda}\,{\sqrt{\rho}}\,\,{\rm arcmin}\\
           &=& 1.2\, \frac{\sqrt{\rho}}{E}\,\,{\rm deg}
\end{eqnarray}

\noindent
where $\rho$ is the mass density of the mirror surface material in gm\,cm$^{-3}$, and $\lambda$ the photon wavelength in Angstroms \cite{attwood00_xrayoptics}. Imaging around 10\,keV requires grazing incidence angles of small fractions of a degree. Bringing these photons to a sharp focus then necessitates long focal lengths. The \nustar\ mission, for instance, uses a deployable mast about 10\,m in length \cite{nustar}. 

Imagine, instead, a lunar X-ray observatory anchored to the surface. This location should allow the requisite space and infrastructure deployment needed for operation and control of long focal length telescopes. A rigid structure, such as a dome or a basic skeletal frame, capable of supporting the mirrors, orientating them, and tracking cosmic sources would be required. The X-ray detectors need not be physically attached to the same structure, if they can be tracking-locked to the mirrors, say. An artist's impression of one such  conceptual schematic design in shown in Fig.\,\ref{fig:xrayobs}. If such ambitious designs can be realised, they would open up cosmological studies of the growth of cosmic structure in the early universe, in addition to a multitude of other science cases beyond the sensitivity and resolution of current facilities. 

An important design consideration here is that mirror effective area will scale down as the projected surface area ($\sim$\,${\rm sin}\,\theta)$ decreases, so shallower incident angles would correspond to smaller effective areas. Detailed engineering simulations will thus be needed to verify the practical limitations on useful focal length enhancements. Size will ultimately be constrained by the mirror and dome body mass that can be transported to the lunar surface. Future concept designs for large-area X-ray mirrors with state-of-the-art focusing optics include mono-crystalline Silicon as a low-density, high tensile strength substrate free of stress. The {\em AXIS} telescope concept, for instance, aims to achieve an order-of-magnitude improvement in effective area over {\em Chandra}, while maintaining similar imaging quality, and a total optics mass of $\approx$\,1\,tonne \cite{axis}. Materials such as a Silicon would also be readily extractable from lunar mines for in-situ manufacturing in the more distant future. 

Similarly, lightweight, stiff materials (e.g. Carbon fiber or other composites) would be preferred for the support structure. Larger payload capacities such as with SpaceX's Starship (promising up to 100\,tonnes of payload capacity\footnote{\url{https://www.spacex.com/vehicles/starship/}}) would facilitate larger, more ambitious designs for lunar installation \cite{elvis_starship}. The absence of lunar winds mitigates the need for a fully covered protective dome, allowing for structure construction to be relatively light, though some additional shielding and collimation to protect the detectors from background X-rays will be required. 

Simpler designs with large array detectors that maximise collecting area without optimising focusing optics have also been considered before \cite{gorenstein90, peterson90}. An array of high throughput collimation detectors would act as excellent `photon buckets' for timing science, and could enable a wide range of transient studies, discussed later in the article. Such designs ought to be amongst the first X-ray pathfinders to be deployed on the lunar surface. 

Another concept that could benefit is the lobster eye mirror design. These mirrors mimic the reflecting tubes that have evolved naturally as part of lobster eye vision, and provide the unique capability of ultra wide field imaging. Such a design was conceptualised several decades ago \cite{angel79}, but only recently demonstrated in orbit with an Einstein Probe pathfinder mission, {\em LEIA}, covering an instantaneous field-of-view of 340\,deg$^2$ \cite{ep_pathfinder}. The typical collecting areas of current designs is modest (a few cm$^2$). Stacking large arrays of microchannel plates is impractical in orbit, but could be done on the Moon to boost the focal collecting area. 

\subsection{X-ray Interferometry}

Normal incidence focusing telescopes can be (and have been) designed to be diffraction-limited at all wavelengths from the ultraviolet to the radio bands. A $d$\,=\,1\,m diameter optical telescope operating at a wavelength $\lambda$\,=\,6000\,\AA, say, yields an effective Airy disc size angular resolution $\theta$\,$\sim$\,0.15\,arcsec. X-ray observations have the potential to deliver enormous gains in angular resolution relative to longer wavelengths, given that the Rayleigh diffraction criterion scales linearly with wavelength $\lambda$. An X-ray telescope sampling 1\,m of an incoming X-ray wavefront at an energy $E$\,=\,2\,keV (say, equivalent to $\lambda$\,=\,6.2\,\AA) should, in principle, be capable of achieving an angular precision three orders of magnitude better ($\theta$\,$\sim$\,0.15\,milli-arcsec). But X-ray telescope designs are not, currently, diffraction-limited because of the lack of technology capable of polishing mirror surfaces that are commensurately smooth to within a fraction of a wavelength. 

An alternative approach would be to utilise interferometry, which still remains to be realised for cosmic X-ray studies. Bringing together signals from multiple telescopes (fringe combining) results in effective gains in angular resolution of $\sim$\,$L$/$d$, where $L$ is the characteristic baseline separation of the array. Sparse interferometry with just a few telescopes primarily yields a complex `visibility' -- a measure of the spatial extent of the target, but better sampling of the image plane over time can be used to reconstruct source images. Interferometry thus allows $\theta$ to improve indefinitely beyond the confines of single telescopes, in principle, being ultimately limited by the sensitivity resulting from the finite collecting area of an array. 

Uttley et al. \cite{uttley20} outline a concept design of an orbiting facility that could leverage X-ray interferometric gains even within a single spacecraft.\footnote{based upon flat reflecting mirror design proposed for the MAXIM mission \cite{maxim}.} Phase interferometry on baselines as short as $\sim$\,20\,m would open up a vast suite of cosmic X-ray source studies, including direct imaging and resolution of the accretion discs around accreting supermassive black holes, the orbits of Galactic X-ray binaries, and the coronae around nearby active stars, to name a few. Expanding this design to a multi-spacecraft constellation would open up studies at angular resolutions on sub-micro-arcsec scales of the `shadows' of massive black holes at the nuclei of many nearby galaxies, measurements of the elongations of gamma-ray burst afterglows at cosmological distances, and allow the capability to resolve exoplanet transits across X-ray emitting stellar discs. For reference, with a 10\,km baseline, observing 10\,keV (1\,\AA) photons yields $\theta$\,$\sim$\,10\,nanoarcsec. Alternate mission concepts also exist, promising the delivery of a step change in X-ray angular resolution (e.g., \cite{mixim}).

But cosmic X-ray interferometry remains to be demonstrated, and there are enormous challenges associated with bringing this to fruition, specifically in enabling path-length equalization at precisions of order nanometers, across $\sim$\,1\,m+ baselines relevant for realistic telescope collecting areas. Highly rigid systems with stable pointing control  will be needed in a single spacecraft. Thermal stability will also be critical (discussed further in \S\,\ref{sec:site}). Some of these challenges are compounded for a constellation where stabilisation is needs to be maintained across the array baseline connecting free flying spacecraft, which may be separated by tens to hundreds of km or more. 

Placing such an interferometer on the Moon would alleviate some of these extreme design challenges, while introducing others. Baselines can be fixed rigid to the ground for multi-telescope arrays across several metres scales, and gradually extended to several hundred metres or more in the long-term. Filling the $u$--$v$ Fourier plane for imaging reconstruction can also be easily done by taking advantage of the rotation of the Moon, as is now standard for Earth-based interferometers. Tidal locking of the Moon means that this rotation will be relatively slow, favouring observations of order $\sim$\,1\,Ms (a duration of about two weeks). 

\revised{Accurate metrology calibration at X-ray wavelengths will be the key challenge in bringing lunar surface interferometry to fruition. This is true for satellites in-orbit as well, but with the added complication on the Moon of having to account for unwanted shifts resulting from human, seismic or thermal activity. Regular calibration and alignment will be essential. The horizon distance for, say, a 3\,m tall telescope structure on the Moon is about 3\,km, which gives a first estimate of the upper-limit to direct line-of-sight baseline lengths that could be calibrated (if using lasers between various array elements, for instance).} Conversely, customisation of baselines (as is the case for the radio Very Large Array on Earth) could also be implemented on-demand for specific angular resolution needs, without the restriction of a limited fuel supply that would be relevant in-orbit.

Additional challenges will include stringent requirements on thermal stability. Specifically, precise beam positioning as well as stability of target staring will require exceptionally low thermal coefficients for the telescope structural components, or will require isothermal controls at $\sim$\,mK levels. \revised{These challenges might be addressed through a combination of judicious site selection, appropriate thermal control technologies, and developing strategies for rapid path length recalibration and realignment. All of these are discussed further in \S\,\ref{sec:site}.}

\section{Astrometry with Occultation studies}

Occultation of a background light source by a foreground body offers the possibility of high angular resolution studies. This relies on accurate {\em timing} in order to enhance spatial resolution, assuming that the instantaneous location of the foreground body is known well enough, and that the occultation time can be measured with sufficient accuracy. With enough signal, such studies can be used to enhance spatial localisation of the background source, study it size and shape, and also search for sub-structure within both bodies. 

In the early days of X-ray astronomy, lunar occultations were utilised as a means to characterise the spatial profile of extended sources. For instance, Bowyer et al. \cite{bowyer64} observed a lunar occultation of the Crab Nebula using an Aerobee rocket, measuring the extent of the nebular emission. Point-source astrometric measurements have also been improved through occultation measurements (e.g., \cite{janes73, davison77}). The need for occultation studies was largely alleviated as focusing X-ray observatories became mainstream, and as their angular resolution improved. Achieving angular resolutions much better than a few arcsec still remains challenging, though, with the few facilities such as \chandra\ capable of high quality imaging being expensive to construct and launch, and with pointed X-ray observations covering only a small field-of-view. \revised{Current surface polishing techniques remain far from the ultimate goal of delivering smooth mirrors capable of diffraction-limited imaging in X-rays. Better precision is possible, and commonly employed, at longer optical wavelengths -- but these are easily obscured and scattered by foreground dust. X-rays offer the possibility to penetrate through heavy obscuration. Thus, X-ray occultation studies could open up new possibilities to study most of the obscured regions of the Milky Way disc, if they can be implemented.}

A return to the Moon offers the possibility of a revival of X-ray occultation studies. This is because large-area detectors can now be made relatively cheaply. A good example here is {\em NICER} \cite{nicer14}, which routinely observes transient sources gathering hundreds of counts per second, but at the expense of spatial resolution. This mitigates a key limitation of past occultation studies, which were photon-starved. Being able to place large-area telescopes on the Moon could then substantially improve on timing and occultation work. 

Below, we consider two kinds of occultation events -- due to natural, and artificial, foreground occulters. The most ambitious of these will be a challenge, but if feasible, could push X-ray studies towards accurate (fractional arcsec) astrometry, a field that has so far been the domain of only longer wavelengths.

\subsection{Natural foreground occultation sources}

The instantaneous projected area of the Earth as seen from the Moon is 2.8\,deg$^2$ on average, about 13 times larger than that of the Moon seen from the Earth. Over the course of a sidereal month (27.3\,days), the Earth would sweep out $\approx$\,660 deg$^2$ of the sky from a lunar vantage point. Due to precession of the lunar orbit, over 18.6\,years the full sweep will encompass an area just shy of 20\,\% of the sky. For a lunar site with a view of the Earth, this should allow a wide range of occultation studies using the Earth as the foreground occulter. 

Absorption and refraction through the Earth's atmosphere will blur the event, reducing astrometric precision, and would also be strongly energy dependent. Solar Wind Charge eXchange (SWCX) radiation from the Earth's magnetosphere and sheath will also create an elevated background at soft energies around the Earth's limb \cite{wargelin04, snowden04, fujimoto07}. A more interesting possibility could then be to instead utilise lunar topography as the foreground occulter. A near equatorial site would see the entire sky rise and then set once a month. All sources in the sky would thus be occulted by the lunar landscape twice every lunar month. 

Here, we explore positional precision in an occultation event. This will depend upon the source (\cs) and background (\cb) count rates, together with the defined signal-to-noise ratio (\snmin) threshold corresponding to a significant detection. We model our estimates follow the algorithm outlined by \cite{mereghetti90}. Let us assume a small time bin \Dt\ during which counts are accumulated. The simplest case is to assume the high count rate limit relevant for bright sources (e.g. Galactic transients), where uncertainties are Gaussian, and with two consecutive time bins straddling the instant of occultation (i.e. the target being visible in one bin and then disappearing in the next). A source can be robustly claimed to have entered occultation within a time interval $\Delta T$ if

\begin{eqnarray}
C_{\rm S}\,\Delta T\, \ge\, SN_{\rm min} \times \sqrt{C_{\rm B}\Delta T+(C_{\rm S}+C_{\rm B})\Delta T}.
\label{eq:mereghetti}
\end{eqnarray}

In the small counts regime, this is modified by the need to incorporate Poisson uncertainties instead. Here, we approximate the overall noise by propagating the confidence limit formulae presented by Gehrels \cite{gehrels86}. A detailed assessment of timing precision will require proper simulations, and will be able to utilise systemic differences in the light curve levels pre-, and then post-, occultation, in order to improve statistics beyond the information provided by the two consecutive bins straddling the occultation. It will also be instrument-specific, dependent on deadtime effects, variability in source flux, and limited by the background. These details are beyond the scope of our evaluations here, which should thus be considered as a baseline upon which to build future simulations. 

The angular speed of the occulter ($\Delta s$/\Dt) projected on the direction of areal motion of the background source (at a relative angle $\alpha$, say) then governs the angular precision $\Delta s$. In this scenario, the angular speed is $\approx$\,0.55 arcsec\,sec$^{-1}$ for both the Earth and for lunar topographic features, as seen from the surface.

\begin{figure}
    \centering
    \includegraphics[angle=0,width=\textwidth]{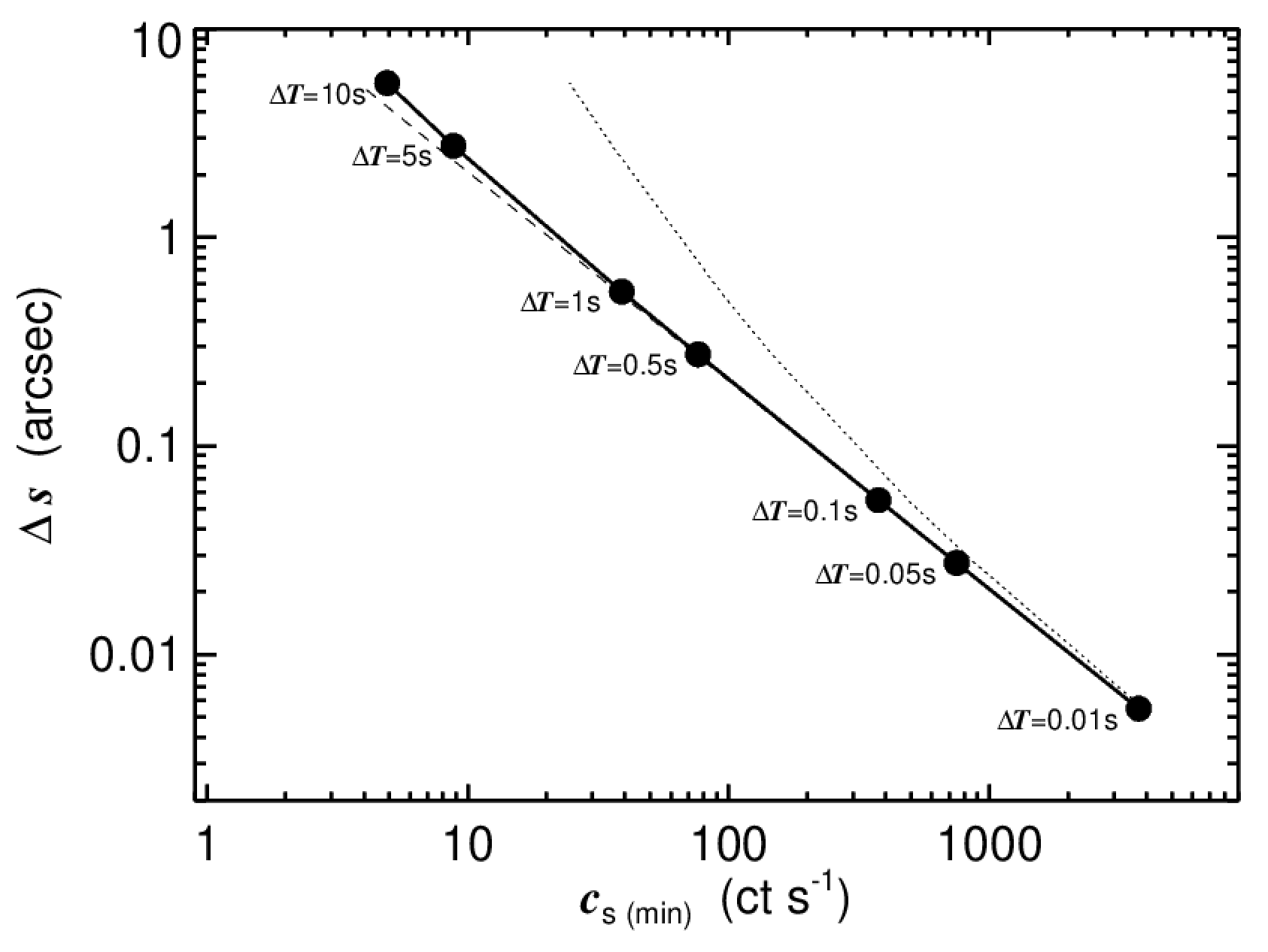}
    \caption{Astrometric constraints possible with occultation studies from the Moon. Localisation precision $\Delta s$ in arcsec is shown as a function of source count rate. The solid curve is the prediction for a background rate of \cb\,=\,1\,ct\,s$^{-1}$. The dotted and dashed lines denote rates of 0.01 and 100\,ct\,s$^{-1}$, respectively. Time bins \Dt\ corresponding to a signal-to-noise detection threshold of 5 are annotated. 
    }
    \label{fig:mereghetti}
\end{figure}

Fig.\,\ref{fig:mereghetti} shows the resultant precision $\Delta s$ possible, for a wide range of count rates, assuming \snmin\,=\,5, and $\alpha$\,=\,90$^\circ$ between the limb of the occulter and the projected areal direction of motion. The solid curve corresponds to \cb\,=\,1\,ct\,s$^{-1}$, whilst the dotted and dashed curves straddling it on either side correspond to background rates two dex higher and lower, respectively. 

The figure demonstrates the impressive gains in spatial localisation that can be made at high count rates. Specifically, at \cs\,$\approx$\,tens of ct\,s$^{-1}$, $\Delta s$ already matches the current best X-ray angular resolution of $\approx$\,0.5\,arcsec delivered by \chandra. At 500\,ct\,s$^{-1}$, $\Delta s$ is a few $\times$\,10\,mas. In this regime, it becomes possible to not only accurately localise and associate source counterparts unambiguously, but also to start to measure their proper motions, directly in X-rays. For reference, 10 ct\,s$^{-1}$ with a telescope of the collecting area of {\em NICER} corresponds to an X-ray flux $F_{0.5-10}$\,=\,2\,$\times$\,10$^{-11}$\,erg\,s$^{-1}$\,cm$^{-2}$ for a typical X-ray power-law spectrum with photon index $\Gamma$\,=\,2.\footnote{Count rate $N_{\rm counts}$\,$\propto$\,$E^{-\Gamma}$.} Larger collimator arrays could push this sensitivity deeper without the need for focusing optics, if the background level can be restricted to a reasonable level. 

Some relevant limitations to consider here include the diffraction of photons around any occulting limb. The relevant Fresnel scale in this case will be \revised{0.6\,$\sqrt{(L/1\,{\rm km})(\lambda/6.2\,{\angstrom})/2}$\,mm for 2\,keV X-ray photons (say), with $L$ being the distance to the occulter, which would lengthen the occultation shadow on the detector, in principle. However, small X-ray wavelengths render diffraction effects much less important than those commonly seen in the radio to optical regimes \cite{roques87}.} Furthermore, physically extended emission regions such as scattering halos surrounding compact accreting sources could lengthen the occultation event or reduce its contrast. Conversely, occultations can be used to find and study the substructures of such extended emission zones (e.g. \cite{mitsuda90}). Finally, $\Delta s$ will ultimately by limited by the precision of local topographic mapping. For a hill at distance $d_{\rm km}$ km from an observing site, the height of the summit will need to be known to a precision better than $\Delta h$\,=\,0.5 ($\Delta s$/0\farcs 1)\,$d_{\rm km}$\,mm. Utilising artificial occultation devices, instead, could alleviate some of this limitation, as discussed below.

\subsection{Customised occultation devices}

Having telescopes on the lunar surface within physical reach enables on-demand customisation of hardware, tailored to specific observations. Artificial occulters could be deployed such that an occultation will occur for a source of interest at specific times and at any altitude and azimuth. These could be simple metallic or other custom-designed opaque plates. Deciding the number, shape, smoothness, and height above ground of the occulter then becomes a (relatively easy) engineering problem to address.

Artificial occulters would be most beneficial for newly discovered transients, when a natural occultation is not imminent. The expected astrometric gain should be similar to that discussed in the previous section, and would circumvent the challenge of high accuracy topographic relief mapping. 

Occultation by a plane is a one-dimensional event, so the positional accuracy is maximised in the direction normal to the occultation, and the calculations above assume $\alpha$\,=\,90$\fdeg$ Arranging for two occultations, with edges at orthogonal angles to each other (e.g. with a suitably oriented wedge) would allow gains to be made in two directions, resulting in a better positional reconstruction using two-dimensional astrometric information. 

\subsection{Astrometric Reference Frames}
Astrometry requires cross-calibration to a reliable coordinate reference frame. For orbiting satellites, default pointing uncertainties can be of order arcsec or larger, unless fine guiding instrumentation is deployed. Even with fine guiding, high astrometric accuracy can be difficult unless the observations utilise instrumentation with large fields of view to encompass multiple background objects within a single observation. Calibration then relies upon tying some of these objects to known astrometric catalogues. 

A fixed observing site on the Moon, together with occulters whose relative location is unambiguously known, mitigates much of the uncertainty associated with pointing inaccuracies. This will allow \revised{generation and cumulative refinement} of an accurate absolute astrometric reference frame relevant to any given site. Regular calibration will undoubtedly be required to account for instrumental and background deviations (discussed further in \S\,\ref{sec:site}). 

\section{Multimessenger Time-domain Coordination}

Opening up new parameter space inevitably leads to new discoveries \cite{harwit}. Multiwavelength and multimessenger studies have demonstrated this repeatedly over the years through a wealth of astrophysical discoveries, because many astrophysical phenomena are inherently broadband sources of radiation. 

Sources that are spatially compact are additionally expected to show {\em time-domain} correlations between various bands. Some of the most extreme changes in time are associated with non-thermal radiation processes such as those found in Gamma-ray bursts, kilonovae, active galactic nuclei, tidal disruption events, magnetars, and X-ray binaries, to name a few examples. Their discovery has often relied on having observing cadences sampling $\sim$\,weekly timescales, with some campaigns able to organise intensive daily monitoring. But physical changes often occur faster than this. These could be state transitions in X-ray binaries, fast radio bursts, periodic eruptions, jet switches (on or off), and more. 

\begin{figure}
    \centering
    \includegraphics[angle=0,width=1.01\textwidth]{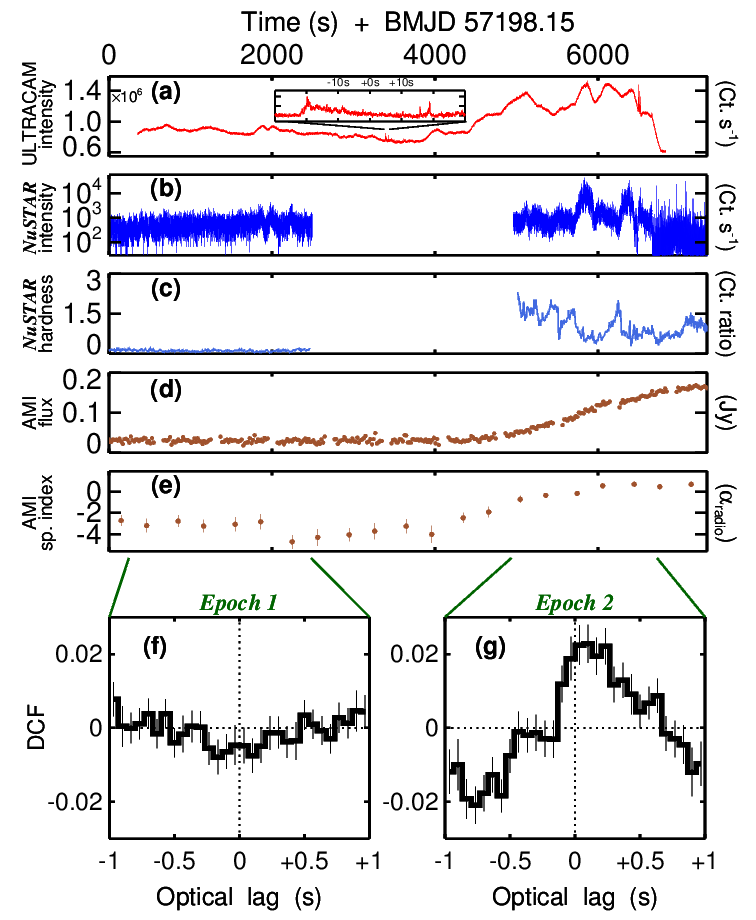}
    \caption{An example of the need for {\em strictly simultaneous} monitoring across the electromagetic spectrum. The black hole X-ray binary V404\,Cyg underwent a spectacular but short-lived accretion outburst in 2015. This figure shows the only set of coordinated optical (Panel a), X-rays (b, c) and radio (d, e) observations that could be arranged at high time resolution with {\em ULTRACAM}, \nustar, and AMI. Other opportunities were missed due to day/night visibility constraints. Even here, Earth occultation of \nustar\ resulted in missing the instant of a physical state transition apparent from the dramatic changes visible between Epochs 1 and 2, separated by the empty gap of X-ray data in the middle. Note the significant rise in X-ray brightness (b) and spectral hardness (c) in Epoch 2, similar effects in the radio (d and e) and the appearance of  significant optical/X-rays timing correlations (DCF; panels f, g), none of which is visible pre-occultation during Epoch 1. The instant of transition may have been associated with the sudden appearance of rapid (sub-second) optical fluctuations visible in the inset in panel (a) during the X-ray data gap. From \cite{g17}.}
    \label{fig:gx339}
\end{figure}

Current monitoring strategies are often too patchy to catch these events, leading to missing out on evidence for some of the key physics connecting the various bands. The issue is compounded when the expected changes span more widely separated bands. One example is shown in Fig.\,\ref{fig:gx339}. This presents a rapid state transition that occurred during the accretion outburst of the Galactic black hole X-ray binary V404\,Cyg. Two X-ray epochs are shown, separated by a gap in the middle caused by one of the regular Earth occultations every $\sim$\,90\,minutes typical for orbiting satellites, when the target is hidden by our planet. The source behaviour differs dramatically between the two epochs, with the second one being significantly brighter and spectrally `harder' (i.e. peaked towards higher energies) in both X-rays and radio (panels b--d). Significant inter-band timing correlations also appear in Epoch 2, whilst they are absent in Epoch 1 (panels f--g). Earth occultation lasted just over 2,000 seconds, but deprived us of the exact instant of transition in X-rays. The source went on to become one of the brightest cosmic sources in the sky some hours later (for details, see \cite{g17}).

This example illustrates some of the challenges associated with current multiwavelength coordination. Observers have to contend with visibility constraints at multiple sites, typical $\sim$\,50\%+ losses due to Earth occultation (relevant for both ground and space telescopes), not to mention time lost as a result of differing weather conditions at different sites. A detailed discussion of the barriers to coordinating observational campaigns can be found in Middleton et al. \cite{smwa}. Additional human factors include the current \lq double jeopardy\rq\ associated with the need to have proposals accepted by  multiple time allocation committees for a single coordinated campaign, and the lack of a coherent approach to multiwavelength proposing that can enable such observations as a matter of course. 

As we push the boundaries of time-domain astrophysics, the demands on the need for strictly coordinated observations will rise. The campaign on V404\,Cyg was far from well planned, and emerged in near real-time as the spectacular nature of the source became clear to astronomers worldwide. Opportunities for coordinated tracking of the outburst evolution were missed as a result. Despite the best will of all interested parties, having widely separated facilities each with their own operational constraints can result in disjointed efforts. 

The Moon offers the possibility of surmounting these barriers, by allowing a single site to host facilities capable of detecting photons across the entire electromagnetic spectrum. Telescopes from radio to X-rays (and beyond) at a single location would all be subject to (near) identical visibility constraints, and free of atmospheric disturbances. This could allow astronomers at all wavelengths to reconsider and optimise coherent approaches to observation planning and delivery. {\em Strict} multiwavelength and multimessenger coordination could become the default mode of operation. 

The long lunar synodic day (29.5 Earth days) would allow lengthy {\em uninterrupted} multiwavelength monitoring of transient and variable events, at least $\sim$\,30\,$\times$ longer than what is possible in Earth (ground and low orbit) observations. This will open up new observation possibilities for transient events in the Fourier frequency regime of $\sim$\,$\mu$Hz--mHz. Dual observatories on the \revised{near} and the \revised{far} sides could remove virtually all sampling gaps, delivering unprecedented coordinated monitoring data. 

Such a multiwavelength facility could also serve as an focus for  international collaborative efforts, if multiple agencies would agree to lead the delivery of infrastructure required in individual bands (for example), with everyone benefitting from the strict simultaneity of the data that are ultimately gathered.

\section{Site Considerations}
\label{sec:site}

Unlike low frequency radio telescopes targetting the early universe, X-ray observatories do not need to be located on the \revised{far} side of the Moon. Power requirements are not expected to be excessive either, at least in the various case studies envisaged herein, though such requirements will inevitably scale in proportion with the size of facility designs. 

Though there will be important challenges to overcome, X-ray observatories could thus benefit from the fact that they will be affected by different (and, in some cases, fewer) site constraints as compared to other facilities. The paucity of `prime' observing sites on the lunar surface (cf. Krolikowski \& Elvis, this volume) due to a combination of human activity and stringent requirements of other wavelengths and messengers is an issue that should not be underestimated. 

Below, we discuss several of the primary challenges and considerations that will need to be addressed for any operating X-ray facility on the lunar surface. 

\subsection{Lunar Dust}

\revised{Lunar dust refers to the fine and loose particulate component of regolith, smaller than pebbles. Dust typically comprises micron-size, abrasive debris left over from impact disintegration of Silicate and Basaltic rocks \cite{mckay91}. Exposed to high energy cosmic radiation and Solar wind, dust particles on the Moon are easily charged, and can be electrostatically levitated above the surface, thus posing a hazard to any surface equipment and habitation \cite{rennilson74}.}

Contending with lunar dust will be an important consideration for all planned lunar facilities. For nested shell X-ray mirrors, accumulation of dust will need to be prevented within the narrow inter-shell gaps. So locations very close to human habitation, mining and launch/landing ports would be best avoided. \revised{Dust screens or shields could be deployed that prevent particulate matter entering the mirrors, while being transparent to X-rays. Depending upon the material used, screens may be opaque to certain energy ranges, especially at soft X-rays. This will need to be factored in to design considerations, but could also help screen out some of the particle-induced background counts from the lunar surface (discussed further below). 

Active removal of dust should be possible electrostatically, by lofting charged dust particles away from surfaces using suitable electrically biased plates or wands. Dust particles could also be actively charged using electron beams or plasma jets, thereby allowing them to be lofted and removed \cite{hirabayashi23}. Such devices may either be fitted on the telescopes, or they could be outsourced to service modules and robots in the longer term.} But lunar regolith could also serve the useful purpose of acting as a shield for any sensitive electronic equipment, with light layers of dust capable of protecting from cosmic rays. 

\revised{Ultimately, calibration and regular cleaning will need to become a matter of routine. Slow accumulation of dust can be accounted for, as long as appropriate calibration data are regularly taken. Once human settlements are more developed, regular downtime for maintenance and cleaning regimens can be implemented.}

\subsection{Seismicity}

\revised{Another important factor to contend with will be seismic activity, for which relatively few studies exist. The Apollo Seismic Network operated for over a decade, and showed that the most common kind of seismic activity results from deep moon quakes, with magnitudes $m_b$\,$<$\,3 and a frequency of $\sim$\,500 events per year \cite{nakamura82, goins81, lammlein77}, on average, driven in large part by tidal interactions with the Earth and the Sun. Shallow quakes can be much more powerful, but occur only a few times per year. Meteoroid impacts are also a intermittent source of seismic activity, and are observed to be strongly clustered in time, possibly associated with meteor shower streams \cite{dorman78}.

Not enough is known about the locations and focal origin points of deep and shallow quakes. These appear to be patchy and focused on the near side according to the data obtained thus far, though this is likely to be a selection effect of the current monitor sampling \cite{nakamura82}. The Chandrayaan-3 Vikram lander, which recently touched down at a southern lunar latitude, carried an instrument to detect seismic activity in these ill-explored zones \cite{sinha23}. In the longer term, we should learn a lot more from ambitious concepts such as the Lunar Gravitational-Wave Antenna, which will comprise a correlated network of seismometers aimed at providing several orders-of-magnitude improvement in sensitivity over the Apollo network in the mHz range \cite{lgwa}. 

Locally strong seismic activity will inevitably interrupt observations. Focusing and pointing recalibration strategies will be important to implement, to ensure speedy realignment following seismic events. Earth-based telescopes now routinely include anti-seismic measures such as shock absorbers that protect sensitive elements and provide some insulation from quake movements \cite{vista}. Daily regimens of pointing calibration are also a standard part of most ground observatory operations. Seismicity is likely to be less of a complication for timing collimators, but will impact science cases with stringent path length tolerances such as a long focal-length telescopes and interferometers. These challenging concepts will likely benefit from more conventional designs (e.g. monolithic, physically connected elements) in early stage designs. The technology for advanced seismic isolation stages and suspension systems now exists, and has been demonstrated in gravitational wave interferometers \cite{matichard15}. Porting some of this technology could help to mitigate environmental noise for lunar X-ray interferometry. 

}

\revised{
\subsection{Thermal stresses}

The lunar surface represents an extreme thermal environment, with temperature variations of more than 250\,K in exposed equatorial surface regions \cite{williams17}. Directed and reflected Solar radiation, as well as reemitted infrared radiation, all act as incident heat loads for any surface body during day time. Conversely night-time operations require added heat retention capacity against the bitterly low surface temperatures. The overall impact of these variations is thermal stress on telescope frames and electronics, resulting in long-term fatigue. 

These are important considerations for X-ray optics, where component tolerances can be more demanding than other wavelengths. Materials with exceptionally low thermal expansion coefficients will be required, in addition to development of novel radiators and thermal insulation. These would also have to be tolerant to dust, which will inevitably accumulate on exposed surfaces over time and change their thermal regulation properties. Conversely, even thin layers of regolith can themselves act as very effective insulators for moderating temperature extremes (variations are known to drop to less than a few deg at shallow depths of just tens of cm \cite{vasavada12}), and so could potentially be incorporated into thermal protection designs of specific telescope and electronic components. This would not solve the problem of thermal stresses on the sky-facing telescope surfaces, of course.

Active cooling strategies adopted in X-ray telescopes now include Transition Edge Sensors.} These are superconducting sensors with exquisite sensitivity to discriminate the energies of incident photons. As a result, they are expected to deliver orders-of-magnitude improvements in spectral resolution with upcoming microcalorimeter arrays (e.g., \cite{xrism, athena, xcalibur}). When cryogenically cooled to $\sim$\,50\,mK, thermal noise is suppressed and results in spectral resolution $E/\Delta E>1,000$ at astrophysically important energies around 6\,keV. 

Lunar observatories would offer the possibility of on-demand coolant refilling, and any cryogenic cooling instruments could also benefit from the mining $^3$He on the Moon. Alternatively, closed cycle coolers are increasingly replacing cryogenic technology. These can generate microphonic noise which not only worsens spectral resolution, but could also potentially destabilise beam combination. Physically separating the cycling coolers from the detectors would help to lessen any microphonic noise, and will be easier to implement on the lunar surface, free from the confines of a spacecraft. 

A side benefit of the high spectral resolution enabled by microcalorimeters is enhanced fringe sampling and a larger field-of-view for X-ray interferometers \cite{uttley20}. This would, in turn, help with interferometric source localisation, overcoming the current deficiency of targets that have known coordinates accurate at the sub-milliarcsec level, a requirement for fringe acquisition and combination. 

\revised{For night time heat retention and operation, radioisotope heater units (RHUs), radiothermal generators (RTGs) and thermal capacitors are some technological solutions being developed. RHUs have been successfully used since the early days of lunar exploration \cite{ulamec10}. RTGs typically incorporate highly toxic and difficult-to-obtain materials with short half lives (e.g. radioactive isotopes of Plutonium and Polonium), and obviously also come with radiation safety concerns. Other elemental isotopes are under active consideration, e.g. Americium-241 \cite{ambrosi19}.}

Ultimately, though, there will be limitations to the surface areas that can be actively cooled (or heated). \revised{For the more challenging science cases still at the proof-of-concept stage (e.g. X-ray interferometry), smaller monolithic structures that can be thermally controlled more efficiently would be the obvious starting point.} Any such design considerations will have to be combined with judicious site selection. Thermal gradients could be mitigated by locating facilities in either the warmer zones of the shadowed lunar regions, or in the high illumination regions (the so-called peaks of eternal light).

\revised{
\subsection{Background}
The lunar surface will act as a source of external background. Particle Induced X-ray Emission (PIXE), scattering and X-ray fluorescence (XRF) from the surface will generate elevated environmental counts. Such emission has been detected in a range of experiments dedicated to studying the lunar albedo and surface chemistry \cite{kamata99, bhardwaj07}. Solar-induced scattering is the primary driver of this background, though extra-solar cosmic rays will also contribute. The sunlit side scatters incident Solar X-rays with an efficiency of $\sim$\,10$^{-5}$, peaking at soft energies  \cite{schmitt91}. The `dark' Moon (i.e. the surface not illuminated by the Sun) is quieter by orders of magnitude, though its emission remains to be characterised in detail \cite{wargelin04}. 

Thus, observations at soft energies and at low, near-horizon altitudes will suffer from scattered surface background during lunar day. For example, topographic occultation observations will need care to screen out the resultant soft scattered X-ray background, or to avoid observations in the direction of sunlit topography. The detectors themselves will benefit from collimators and shields (excluded from the schematic impression in Fig.\,\ref{fig:xrayobs} for clarity). Judicious site selection may also help here, if appropriate shadowed mountain sides or craters can be identified as observatory sites. XRF can also fluoresce artificial materials, albeit concentrated in specific transitions \cite{adler77}, which custom-made occulters will need to account for in their designs.

Additional weaker, non-negligible sources of background will include X-ray scattering by the thin dusty plasma that has been elevated in Moon's exosphere \cite{rennilson74}, and charge exchange during passages during the Earth's magnetotail on a few days near full Moon \cite{Narendranath11}. Other natural sky background rates should not differ substantially from the rates seen by telescopes in {\em high} Earth orbit, and will benefit from the absence of the excess noise experienced in Earth's South Atlantic Anomaly. Knowledge of the background impacting other X-ray facilities operating far from the Earth (e.g. at L2) would also be informative. Detailed quantification of these background effects will be important for full characterisation of the overall environmental noise. 

}

\vspace*{0.75cm}
\noindent
Beyond the issues discussed above, facility siting will be dependent upon the desired science goals and calibration requirements.

Galactic astronomy has a long heritage in X-rays. The brightest X-ray sources in the sky beyond the Solar system will be Galactic point sources, mostly accreting compact objects. These would be the best targets for the astrometric occultation science case, as well as the coordination science case, discussed above. An optimal site would thus allow good visibility of the Galactic bulge and centre region. Such a site would still undoubtedly be able to deliver cosmological studies. 

Some of the science cases considered above may best be undertaken from craters with floors a few tens to hundred metres wide, or larger (e.g. to optimise line-of-sight visibility of other unit telescopes in an interferometric array for calibration purposes, or clear sightlines to foreground occultation sources). 

\revised{A constant power supply will need to be maintained on-site. As the efficiency and capacity of industrial-scale rechargeable batteries improve, these could plausibly be stacked together and deployed for lunar night operations of low-power facilities. Monthly battery replacement, or recharging during lunar day using solar panels (cf. Fig.\,\ref{fig:xrayobs}), could be integrated into facility logistical plans. Power demands will scale with the size and `complexity' of the science cases discussed above, with the more ambitious plans likely requiring the presence of developed human infrastructure to sustain their power needs.}

Calibration sources could plausibly be embedded in rocks and topography surrounding the site,\footnote{Putting in place appropriate safety measures for human presence, of course.} providing bright emission sources, free of Doppler variations, at a variety of focal lengths. In the longer term, construction, testing, and calibration of large-scale X-ray telescopes could be moved entirely to the lunar surface, because the absence of an atmosphere removes the need for specialised evacuated beamlines as currently required on Earth. 

\section{Conclusion}

X-ray astronomical observations can only be conducted from beyond the Earth's atmosphere. The benefits of the Moon in this regard may not be immediately obvious, \revised{but our nearest cosmic neighbour has always had a tantalizing relationship with X-ray astronomy. After all, the very first rocket flight that opened up cosmic X-ray astronomy was, supposedly, aimed at searching for X-rays from the Moon \cite{giacconi62}. And the Moon enabled some of the first detailed cosmic occultation studies in X-rays.}

The field has moved on a long way since then, but here we have presented several science cases and engineering considerations that could reap benefits from humanity's inevitable return to the Moon, including reengineering designs to surmount current practical restrictions to telescope focal lengths and effective areas, development of X-ray interferometery, a revival of occultation studies for improving astrometric precision, and developing a coordinated approach to multimessenger time-domain observations. 

\revised{Operating on the lunar surface will also introduce significant new challenges, including elevated background, seismicity and thermal variations, which should not be underestimated.} And ultimately, cost considerations will likely be the governing factor for adoption of any of these proposals. Several of the cases above can be implemented without payload capacity being a restriction. For example, {\em NICER} has a compact and light-weight design, with a launch mass of only 372\,kg.\footnote{\url{https://www.nasa.gov/sites/default/files/atoms/files/spacex_crs-11_mission_overview.pdf}} With Starship promising a payload transport capability of $\sim$\,100\,tonnes, deployment of telescopes with effective area mass tens of times larger than {\em NICER} should not pose a critical limitation. 

Some of the ambitious design infrastructure (e.g. the dome from Fig.\,\ref{fig:xrayobs}) would benefit most from enhanced launch capabilities, and these likely need to await advanced settlement of Moon bases. In the mean time, it should be feasible to deploy lightweight pathfinders that can operate autonomously from the surface, once initial lunar support capability becomes operational, via the Gateway orbiting spaceport and leveraging the proposed regular landings of the European Argonaut missions (Carpenter, this volume).

\ack{PG acknowledges support from UKRI, and thanks the organisers for the opportunity to contribute this article. He also thanks R.\,L.\,Kelley, M.\,Elvis for very useful discussions\revised{, and S. Mereghetti, C. Dashwood Brown, D. Buisson and J. Tomsick for supportive comments and suggestions. Feedback from two anonymous reviwers helped to strengthen the manuscript substantially}.}

%%%%%%%%%% Insert bibliography here %%%%%%%%%%%%%%

\bibliographystyle{RS}
\bibliography{gandhi23}

\end{document}